\documentclass[prb,preprint,showpacs,epsfig,longeq]{revtex4-1}
\usepackage[utf8x]{inputenc}
\usepackage{graphicx}
\usepackage{float}
\usepackage{dcolumn}
\usepackage{bm}
\usepackage{amssymb}
\usepackage{epsfig}
\usepackage{psfrag}
\usepackage{amsfonts}
\usepackage{lineno,hyperref}
\usepackage{color}

\begin{document}

\title[Pressure-dependent Semiconductor to Semimetal and Lifshitz transitions in 2H-MoTe$_2$:]{Pressure-dependent Semiconductor to Semimetal and Lifshitz transitions in 2H-MoTe$_2$: Raman and First-principles studies}

\author{Achintya Bera$^1$, Anjali Singh$^{2}$, D V S Muthu$^1$, U V Waghmare$^2$ and A K Sood$^1$\footnote{electronic
mail:asood@physics.iisc.ernet.in}}

\affiliation{$^1$Department of Physics, Indian Institute of Science, Bangalore 560 012, India}
\affiliation{$^2$ Theoretical Sciences Unit, Jawaharlal Nehru Centre for Advanced Scientific Research, Jakkur, Bangalore 560 064, India}

\date{\today}


\begin{abstract}

High pressure Raman spectroscopy of bulk 2H-MoTe$_2$ upto $\sim$ 29 GPa is shown to reveal two phase transitions (at $\sim$  6 and 16.5 GPa), which are analyzed using first-principles density functional theoretical calculations. The transition at 6 GPa is marked by changes in the pressure coefficients of A$_{1g}$ and E$^{1}_{2g}$ Raman mode frequencies as well as in their relative intensity. Our calculations show that this is an isostructural semiconductor to a semimetal transition. The transition at $\sim$ 16.5 GPa is identified with the changes in linewidths of the Raman modes as well as in the pressure coefficients of their frequencies. Our theoretical analysis clearly shows that the structure remains the same upto 30 GPa. However, the topology of the Fermi-surface evolves as a function of pressure, and abrupt appearance of electron and hole pockets at P $\sim$ 20 GPa marks a Lifshitz transition. 

\end{abstract}

\maketitle
\section{Introduction}
\label{sec:introduction}

In recent years the two-dimensional transition metal dichalcogenide (TMD) materials \cite{bc1} have attracted a lot of attention from the viewpoint of layer-dependent band gap engineering, transistor on-off ratio, high carrier mobilities, effect of spin-orbit interactions, and spin-valleytronic devices. TMDs are promising candidates in opto-electronic applications as they become direct band gap semiconductor \cite{bg1,bg2,bg3,bg4,bg5,bg6,bg7} in mono- and bi-layer limit \cite{bg8,bg9} with a gap in the range of 1.0 to 2.0 eV. Reduced screening of long range Coulomb interactions in the monolayer limit of TMDs results in formation of  of neutral (exciton) and charged (positive and negative) bound quasi-particles (trions) \cite{trion3,trion2} even at room temperature.

One of the unique properties of the group VI TMDs is that they can exist in different structural forms with different electronic properties: (i) semiconducting phase (H-phase), where the metal ions share the coordination with six chalcogen atoms in prismatic configuration; (ii) metallic phase (T-phase) having octahedral coordination, (iii) 1T$^{'}$-phase - a distortion of 1T-polytype structure caused by the anisotropic metal-metal bonding, and (iv) a rhombohedral polytype (3R) phase observed in multilayer MoS$_2$ exhibiting valley-dependent photoluminescence \cite{suzu1} and in Ta$_{1-x}$Mo$_{x}$Se$_2$ showing superconductivity \cite{luo1}. In recent years, the 1T-phases of MoS$_2$ has been synthesized by the intercalation of ions \cite{manish1,manish2} to produce the transparent electrodes and energy storage devices. Recent experiments \cite{dong1} on bulk 1T$^{'}$-MoTe$_2$ show a large mobility $\sim$ 4,000 cm$^{2}$V$^{-1}$s$^{-1}$ and magneto-resistance $\sim$ 16,000 $\%$ at 1.8K in a field of 14 T. A proposal has been made to study the $\mathbb{Z}_2$-topological quantum devices \cite{qian1} in few layer 1T$^{'}$-MoTe$_2$ with a bandgap \cite{dong1} of $\sim$ 60 meV. Recently, high temperature metastable 3R-phase of MoTe$_2$ has been observed with the rich Nb doping \cite{rphase1}.

The electronic topological transition \cite{PRL1,Ett1} or Lifshitz \cite{lif1} transition occurs when the van Hove singularity associated with the band extrema approaches the Fermi level and passes through it, thereby distributing the carriers and hence changing the Fermi surface topology. The subtle changes in the electronic band topology or Fermi surface topology driven by external parameters, which can be reflected in the anomalies of the measurable quantities, are mostly of two types: (a) the appearance or disappearance of electron and hole pockets and (b) the rupturing of necks connecting Fermi-arcs. In our present work, we observe Lifshitz transition due to (a). Raman spectroscopic studies have been shown to be useful to capture the phonon signatures of the subtle modifications in the Fermi surface topology through the changes in pressure coefficients \cite{mgb1} or the integrated area ratios \cite{mgb2} of the Raman modes.

Mechanical strain or deformation is one of the routes for the TMDs to switch among different thermodynamically stable structural polytypes without introducing impurities. Recent high pressure Raman studies \cite{chi,bandaru,nayak1} on 2H-MoS$_2$ reveal that there is a onset of lateral shift of the adjacent S$-$Mo$-$S layers around $\sim$ 20 GPa leading to a mixed phase of 2H$_c$ (2H) and 2H$_a$ structures with the 2H$_c$-phase being the dominant one \cite{bandaru} and thereby changes the pressure coefficients of the Raman modes \cite{chi}. The completion of the layer sliding transition from 2H$_c$ (semiconductor) to 2H$_a$ (metal) occurs at $\sim$ 40 GPa \cite{chi}. Recently, it has been shown \cite{Zheng2} that the pressure-induced metallic transition in MoX$_2$ (X=S, Se and Te) is attributed to the strong coupling of layers and is not  due to the structural transition from 2H$_c$ to 2H$_a$. In this work, we report Raman studies of bulk 2H-MoTe$_2$ as a function of pressure. We see an observable change in the pressure coefficients of the frequencies, S=d$\omega$/dP, of first order $E^{1}_{2g}$, $A_{1g}$ phonon modes at $\sim$ 6 GPa. As shown by the first-principles calculations, this transition is associated with the indirect band gap closing at K$_1$-point, K$_2$-point and K-point in the Brillouin zone (BZ), thereby marking this transition from semiconductor to semimetal. We also observe a decrease of pressure coefficient of the $A_{1g}$ mode from 2.2 to 1.7 cm$^{-1}$/GPa around $\sim$ 16.5 GPa. Our first-principles calculations confirm that there is no structural transition in 2H-MoTe$_2$ throughout the whole pressure range upto 30 GPa. This is consistent with our Raman experiments where no new Raman modes expected for a lower symmetry structure \cite{kan1}. Our findings is quite similar to the cousin material MoSe$_2$, where no structural transition is observed upto the maximum pressure of $\sim$ 60 GPa and it undergoes metallization around 41 GPa \cite{zhao1}.

\section{Experimental details}
\label{sec:experimental Details}

Thin platelets (thickness $\sim$ 50 $\mu$m) cleaved from 2H-MoTe$_2$ single crystals were placed together with a ruby chip inside a hole of diameter $\sim$ 200 $\mu$m in a stainless steel gasket inserted between the diamonds (of culet size 600 $\mu$m) of a Mao-Bell-type diamond anvil cell (DAC). We could not get the reliable Raman data using the Methanol-ethanol (4:1) pressure-transmitting medium. This could be due to the adsorption on the surface of TMDs and the dissociation of alcohols through thermodynamically favorable channels as predicted using first-principles calculations \cite{chen1}. Potassium bromide (KBr) was used as the pressure transmitting medium \cite{kbr1}, the pressure determined via the ruby fluorescence shift. We recognize that in KBr pressure transmitting medium, the pressure is quasihydrostatic above 3-5 GPa \cite{kbr1}. We came across a recent high pressure studies on semiconductor SnSe \cite{quahydro1} where it has been explicitly commented that the x-ray diffraction experiments on semiconductor to semimetallic transition at $\sim$ 12 GPa showed similar observations with (methanol-ethanol-water mixture) and without the pressure transmitting medium and hence we believe that the results presented in this study are generic. Unpolarized Raman spectra were recorded in backscattering geometry using 514 nm excitation from an Ar$^+$ ion laser (Coherent Innova 300). The spectra were collected using DILOR XY Raman spectrometer coupled to a liquid nitrogen cooled charged coupled device (CCD 3000 Jobin Yvon-SPEX). After each Raman measurement, calibration spectra of a Ne lamp were recorded to correct for small drift, if any, in the energy calibration of the spectrometer. Laser power ($<$ 5 mW) was held low enough to avoid heating of the sample and the spectra were collected at each pressure for 15 minutes. The peak positions were determined by fitting Lorentzian line shapes with an appropriate background.\\

\section{Computational details}

Our first-principles calculations are based on density functional theory as implemented in Quantum ESPRESSO package \cite{qe}, in which the interaction between ionic core and valence electrons is modelled by norm-conserving pseudopotentials \cite{SG, CH}. The exchange-correlation energy of electrons is treated within a Local Density Approximation (LDA) with a functional form parametrized by Perdew-Zunger \cite{PZ}. We use an energy cutoff of 80 Ry to truncate the plane wave basis used in representing Kohn-Sham wave functions, and energy cutoff of 320 Ry for the basis set to represent charge density. Self-consistent solution to the Kohn-Sham equations was obtained until the total energy converges numerically to less than 10$^{-8}$ Ry. Structures are relaxed to minimize the energy till the magnitude of Hellman-Feynman force on each atom is less than 0.001 Ry/bohr. We include van der Waals (vdW) interaction with the parametrization given in Grimme scheme \cite{Grimme}. In self-consistent Kohn-Sham (KS) calculations of configurations of bulk 2H-MoTe$_2$ unit cell, the Brillouin zone (BZ) integrations are sampled on 12x12x3 and 24x24x6 uniform meshes of k-points in determination of total energy and electron-phonon coupling respectively. For bulk 2H-MoTe$_2$ at zero pressure, we determine electronic structure by including the spin-orbit coupling (SOC) through use of relativistic pseudopotentials using a second variational procedure \cite{Corso}. Phonon and dynamical matrices at $\Gamma$-point (q= (0, 0, 0)) as a function of lattice constant (or pressure) were determined using density functional linear response as implemented in Quantum ESPRESSO(QE) \cite{qe}, which employs the Green's function method to avoid explicit use of unoccupied Khon-Sham states. Since DFT typically underestimates the bandgap, we have used HSE functional as implemented in QE to estimate the gaps accurately. The mixing parameter is equal to 0.15 in these calculations.  The reciprocal space integration is performed using 108 k-points in each direction for a 6x6x3 uniform mesh.  
\section{Results and Discussion}
\label{sec:Results and Discussion}

\begin{figure}[h!]
	\centering
	\includegraphics[trim= 10 10 15 -10, scale=0.9]{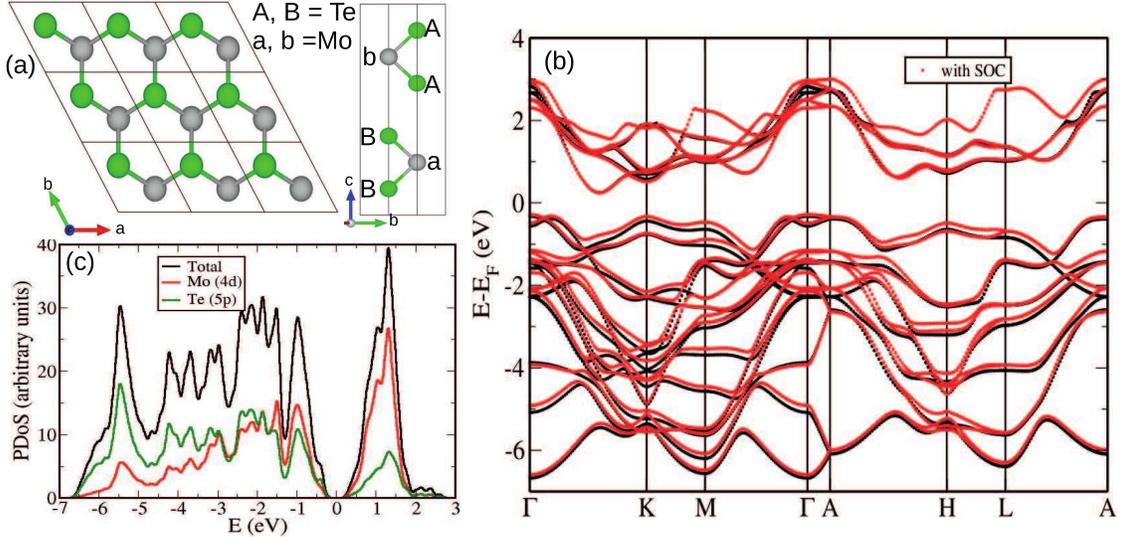}
	\caption{(color online)-- (a) Crystal structure, (b) electronic structure and (c) projected density of states of 2H-MoTe$_2$. Electronic structure determined with (red color lines) and without (black color lines) effects of the spin-orbit coupling (SOC); the effect of SOC are particularly evident in the states lining the gap. }
	\label{Fig_1}
\end{figure}

The structure of 2H-MoTe$_2$ consists of layers Te-Mo-Te, with a unit cell characterized by a stacking sequence AbABaB, where A, B label Te atomic layers and a, b label Mo atomic layers with triangular lattices (see Fig.~\ref{Fig_1}a). 2H-MoTe$_2$ is an indirect band gap semiconductor with valence band maximum (VBM) at $\Gamma$ and conduction band minimum (CBM) at K$_2$ point (K$_2$ point is between the $\Gamma$-K direction) with a gap 0.57 eV (refer to Fig.~\ref{Fig_1}b). The VBM and CBM have contributions from 4d orbital of Mo and 5p orbital of Te (see Fig.~\ref{Fig_1}c). Inclusion of SOC reduces the indirect gap by 30 meV. The splitting of bands due to SOC is relatively smaller at $\Gamma$ point than at K and M points. Since the change in band gap due to inclusion of SOC is small, we have not included SOC in further pressure-dependent calculations.

\par
\begin{figure}[h!]
	\centering
	\includegraphics[trim=10 0 15 20, scale=0.7]{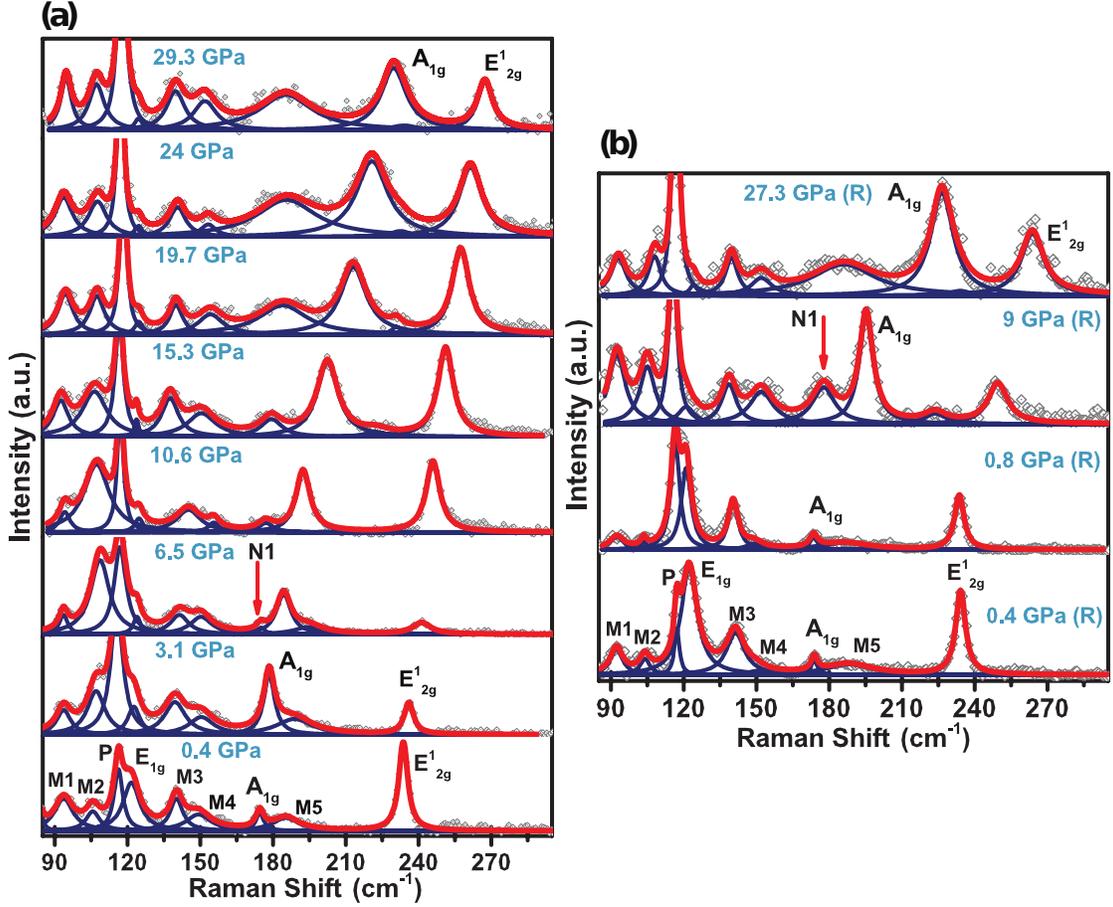}
	\caption{(color online) -- Evolution of Raman spectra with pressure (a) in forward cycle and (b) in return cycle, indicated by R next to the pressure value. Solid lines are Lorentzian fits to the experimental data points (open circles). N1 represents the appearance of a new mode and P stands for plasma line.}
	\label{Fig_2}
\end{figure}

 The space group of 2H-MoTe$_2$ is D$_{6h}^{4}$ with unit cell containing two formula units. The optically active modes at the Brillouin zone centre ($\Gamma$ point) are classified into following irreducible representations as $A_{1g}+A_{2u}+B_{1u}+2B_{2g}+E_{1g}+E_{1u}+2E_{2g}+E_{2u}$. Out of these A$_{2u}$ and E$_{1u}$ modes are infrared active whereas A$_{1g}$, E$_{1g}$ and E$_{2g}$ modes are Raman active. Fig.~\ref{Fig_2}(a) shows Raman spectra of MoTe$_2$ at a few representative elevated pressures inside the DAC. In addition to the Raman active modes $A_{1g}$ at $\sim$ 174 cm$^{-1}$, $E^{1}_{2g}$ at $\sim$ 234 cm$^{-1}$ and $E_{1g}$ at $\sim$ 121 cm$^{-1}$, second-order modes marked as M3 ($\sim$ 140 cm$^{-1}$) and M5 ($\sim$ 185 cm$^{-1}$) are observed which were assigned \cite{saito1} as 2TA[M] or $E^{1}_{2g}$[M]-LA[M] and $E^{1}_{2g}$[M]-TA[M], respectively. Here TA and LA imply transverse and longitudinal acoustic modes and M represents the high symmetry point in BZ. The mode labeled M1 at $\sim$ 94 cm$^{-1}$ can be a combination mode and M2 at $\sim$ 106 cm$^{-1}$ could be disorder activated LA(M) mode \cite{saito1}. Around $\sim$ 6.5 GPa, a new mode (N1) appears at $\sim$ 175 cm$^{-1}$ and continues to evolve with the pressure coefficient of 0.6 cm$^{-1}$/GPa upto the maximum pressure of 29 GPa. As our calculations do not reveal any structural change or the symmetry lowering in the whole pressure range, the origin of this new mode (N1) is not clear. It can be either due to the strain induced splitting of E-type modes or the infrared inactive \cite{yama1}, out of plane mode B$_{1u}$ ($\sim$ 176 cm$^{-1}$) becoming Raman active or it can be related to the $A_{1g}$ ($A_{g}$) mode of T ( T$^{'}$)-phase at 160 cm$^{-1}$ \cite{kan1}. The later possibility is unlikely, because the laser power ($\leq$ 5 mW) might not be able to heat the sample and our theoretical calculations rule out the phase transformations under adiabatic condition. Moreover, the mode N1 is present at different points on the sample above 6 GPa. Further, the mode N1 vanishes in the return-pressure cycle below 7 GPa (shown in Fig.~\ref{Fig_2}b). At this stage, we can only speculate that the mode N1 could be the optically inactive B$_{1u}$ mode becoming Raman active in the semimetallic phase. Another mode M4 is observed at $\sim$ 150 cm$^{-1}$, and we suggest it to be a second order mode of the KBr medium \cite{kbr2}.

\begin{figure}[h!]
	\centering
	\includegraphics[trim=40 0 15 -30, scale=0.6]{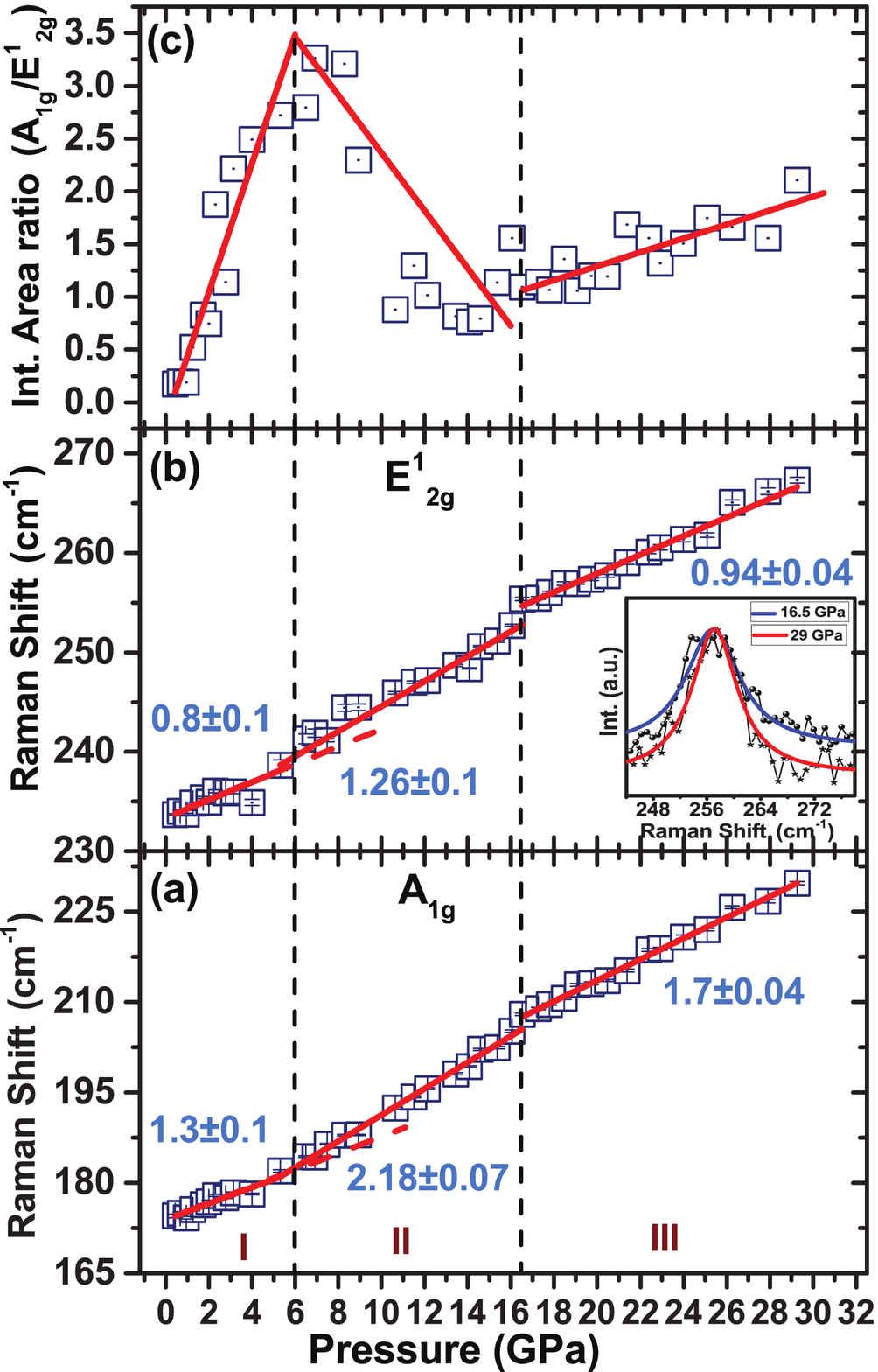}
	\caption{(color online) -- The phonon frequencies of (a) $A_{1g}$, (b) $E^{1}_{2g}$ and (c) the integrated area ratio of $A_{1g}$ to $E^{1}_{2g}$ versus pressure plot. The solid lines are linear fits [$\omega_p=\omega_0+(\frac{d\omega}{dP})P$] to the observed frequencies (solid symbols) and the corresponding slope values are shown. The inset of (b) shows Raman spectra at P = 16.5 and 29 GPa, where the spectra are laterally shifted to match the frequency and also normalized . Error bars (obtained from the fitting procedure) are also shown. The black dashed lines mark the phase transitions and the red dashed lines are guide to the eye.}
	\label{Fig_4}
\end{figure}

At $\sim$ 6 GPa, there is a change in S (=d$\omega$/dP) (Figs.~\ref{Fig_4}a and b) of the phonon modes $A_{1g}$ and $E^{1}_{2g}$. Our first-principles calculations show that this transition is associated with the semiconductor to semi-metal transition (SMT). The change in S across the transition pressure for the out-of-plane $A_{1g}$ mode (0.9 cm$^{-1}$/GPa) is about twice that of the in-plane $E^{1}_{2g}$ mode (0.4 cm$^{-1}$/GPa). We also observe a maximum in the integrated area ratio of the $A_{1g}$ mode to the $E^{1}_{2g}$ mode around 6 GPa (see Fig.~\ref{Fig_4}c) in agreement with our calculations of Raman tensor, to be discussed. 

Fig.~\ref{Fig_4}(c) also suggests another transition at $\sim$ 16.5 GPa. Across this pressure range, the value of S for the $A_{1g}$ mode and the $E^{1}_{2g}$ mode decreases. The changes in S across this transition for both the $A_{1g}$ and the $E^{1}_{2g}$ Raman modes are same (0.5 cm$^{-1}$/GPa). The inset of Fig.~\ref{Fig_4}(b) represents the two spectra at 16.5 GPa and 29 GPa to show a reduction in the linewidth of the $E^{1}_{2g}$ Raman mode at higher pressures. We later interpret this transition to be a Lifshitz transition. The non-hydrostatic condition of KBr pressure transmitting medium would be reflected in the increment of linewidth for the Raman modes as a function of pressure. But on contrary, we have observed the decrement in the linewidth and it is the competitive effect from both the contributions i.e. the non-hydrostatic as well as the Lifshitz transition and it is the latter playing the dominant role in the resultant linewidth. On the other hand, our first-principle calculations not only show under hydrostatic condition that there is no structural transition through out the whole pressure range and there is a onset of Lifshitz transition around 20 GPa which matches well with the experimental observation of transition regime, but also confirm both the transitions (semiconductor to semimetal and the Lifshitz transition) under non-hydrostatic condition as well with a little shift in the transition pressure.

\begin{figure}[h!]
	\centering
	\includegraphics[trim=50 0 15 -20, scale=0.8]{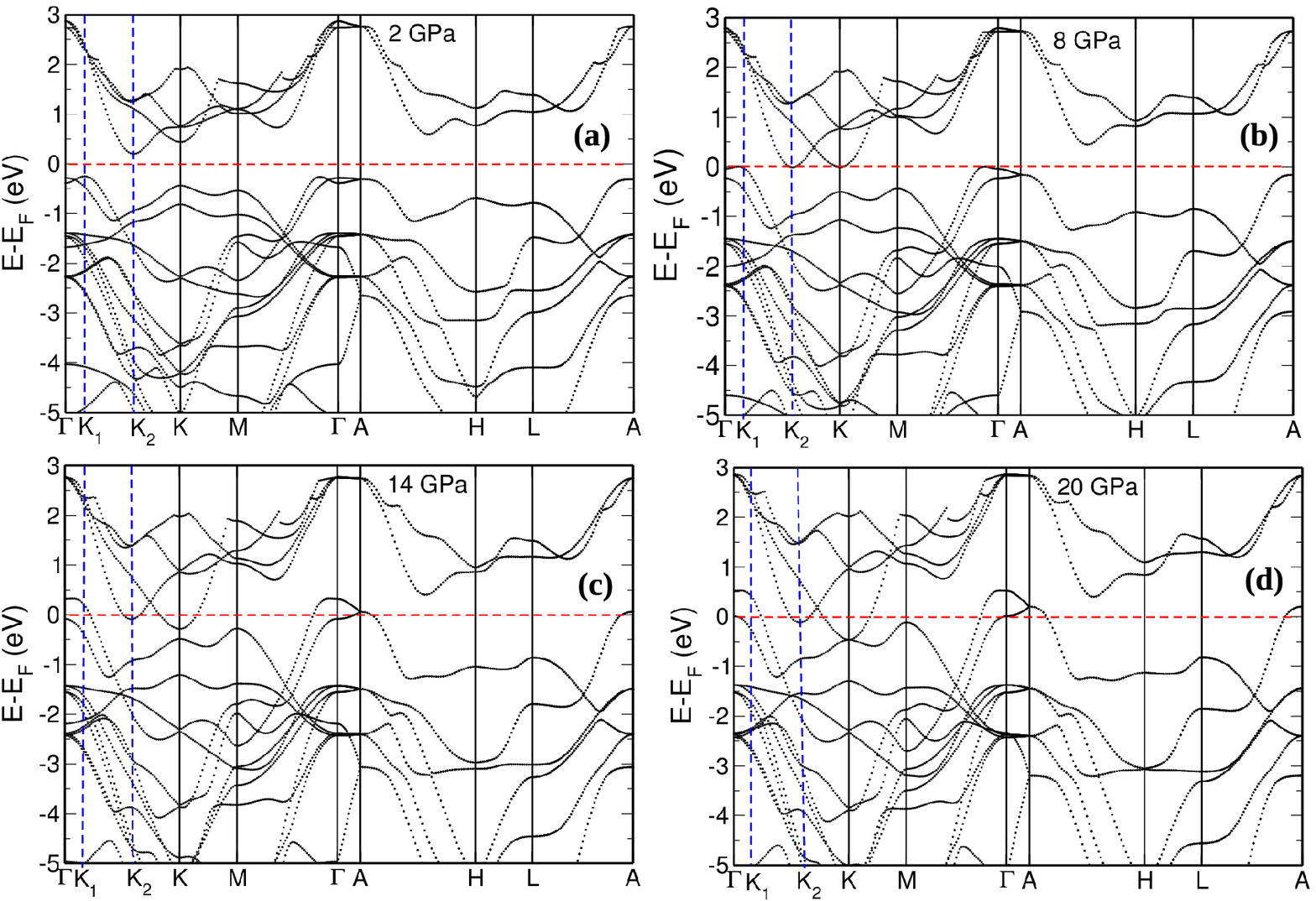}
	\caption{(color online)--Electronic structure of 2H-MoTe$_2$ at (a) 2 GPa, (b) 8 GPa, (c) 14 GPa, (d) 20 GPa .}
	\label{Fig_5}
\end{figure}

\par
\begin{figure}[h!]
	\centering
	\includegraphics[trim=20 0 15 -30, scale= 0.8]{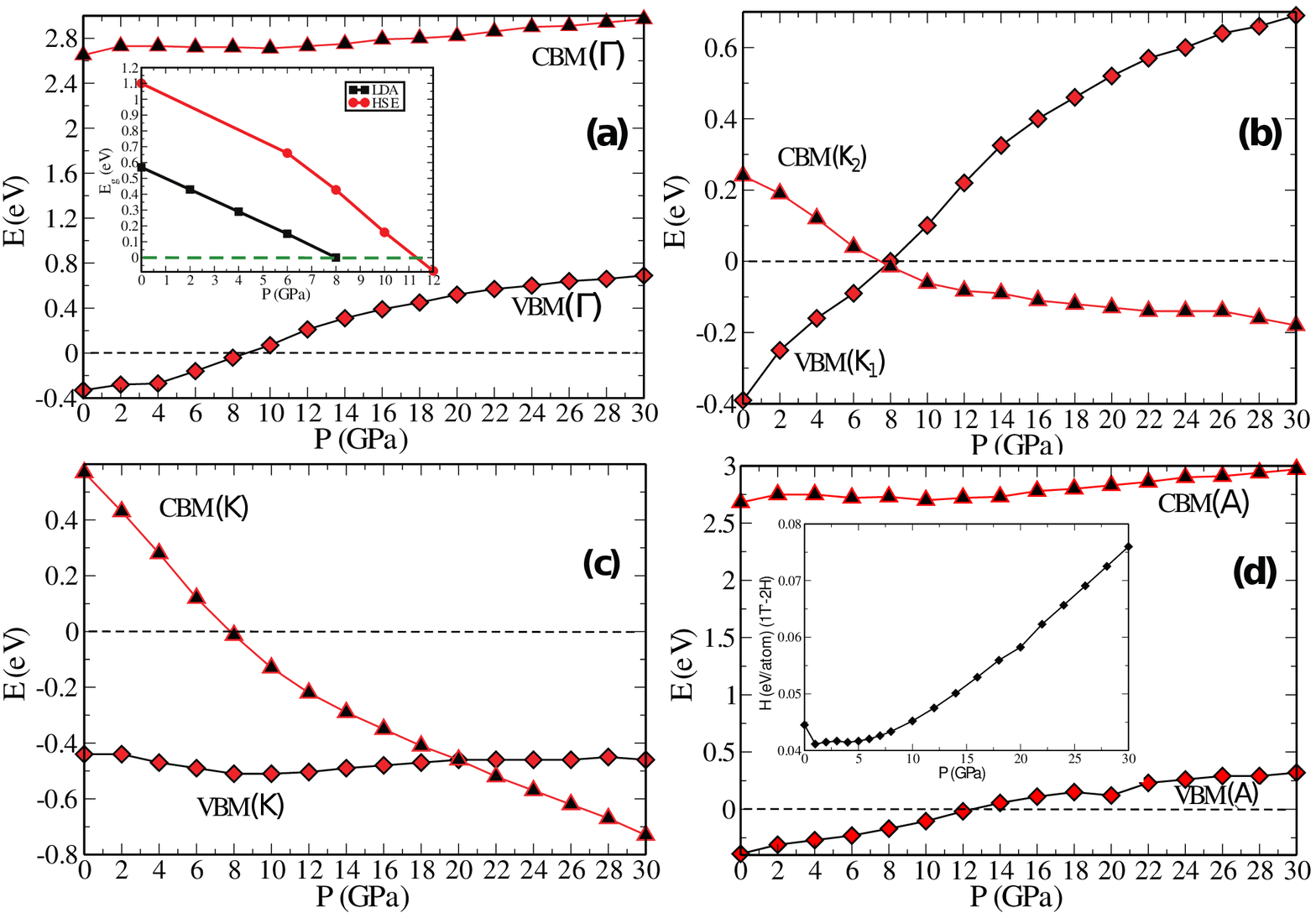}
	\caption{(color online)-- Variation in VBM and CBM with pressure at different high symmetry points of Brillouin zone (a) $\Gamma$, (b) K$_1$ (VBM; near $\Gamma$ point), K$_2$ (CBM; mid point of $\Gamma$-K path), (c) K and (d) A point. The inset in (a) shows the estimates of band gaps with HSE and LDA functionals with pressure. The difference in enthalpy of 2H and 1T$^{'}$-MoTe$_2$ with pressure is shown in inset of (d). Note that 2H-MoTe$_2$ stability increases with pressure.}
	\label{Fig_6}
\end{figure}

 We will now present theoretical calculations to understand the two transitions at 6 GPa and 16.5 GPa observed in the experiments. We find that the structure of 2H-MoTe$_2$ remains stable upto the highest pressure, but its indirect band gap reduces and VBM shifts to K$_1$ point (q= (1/14, 1/14, 0), near $\Gamma$-point) (see Fig.~\ref{Fig_5}a).  We clearly observe (see Fig.~\ref{Fig_6}b) that the VBM (at K$_1$-point) and the CBM (at K$_2$-points) cross the Fermi level at 8 GPa. In addition, the CBM at K-point and VBM at $\Gamma$-point also cross Fermi level (see Figs.~\ref{Fig_5}b,~\ref{Fig_6}a and~\ref{Fig_6}c). At 8 GPa, there are very few states at the Fermi level and hence MoTe$_2$ is semi-metallic, in agreement with the previous calculations \cite{tosatti1}.  At 8 GPa, the indirect band gap with VBM at K$_1$ and CBM at K$_2$ (mid point of $\Gamma-K$ path) as well as at K-point (see Fig.~\ref{Fig_5}b) closes. We note that Riflikov\'a et al \cite{tosatti1} have predicted semiconductor to metallic transition in between 10 to 13 GPa. The inclusion of van der Waal's interaction in our calculations perhaps results in reduction of the predicted transition pressure. The transition observed in our experiment at $\sim$ 6 GPa which is identified as a semiconductor to semimetal transition, is seen to occur at 8 GPa in our calculations. This difference between the observed and calculated transition pressures is partly due to errors in the calculated equilibrium lattice constants. The pressure uncertainties in our experiment and theory are $\sim$ $\pm$ 0.5 GPa and $\sim$ $\pm$ 1 GPa, respectively.

\par Furthermore, we determine the band gap of MoTe$_2$ using HSE calculations to estimate the accurate transition pressure of semiconductor to semimetal transition. The HSE based estimates of the band gap at 0 GPa is 1.1 eV, while the gap estimated with LDA is 0.57 eV. We note that the former is in good agreement with experiment at P = 0 GPa, whereas the latter is an underestimate by 0.4 eV, with respect to the experimental value (1.1 eV). Thus, the pressure of semiconductor to semimetal transition is expected to be underestimated with LDA calculation. $P_c$ obtained with HSE calculation is 12 GPa (see inset of Fig.~\ref{Fig_6}a). However, this is expected to be also off-set by the errors in the lattice constants calculated by LDA, and a precise comparison of these results with experiment on this anisotropic material is tricky. The motivation behind our calculations is to understand the nature of transition rather than predicting the precise transition pressure, and the link demonstrated between the Raman anomaly and the electronic transition is physically reasonable.

\begin{figure}[h!]
	\centering
	\includegraphics[trim=0 0 15 -30, scale=0.8]{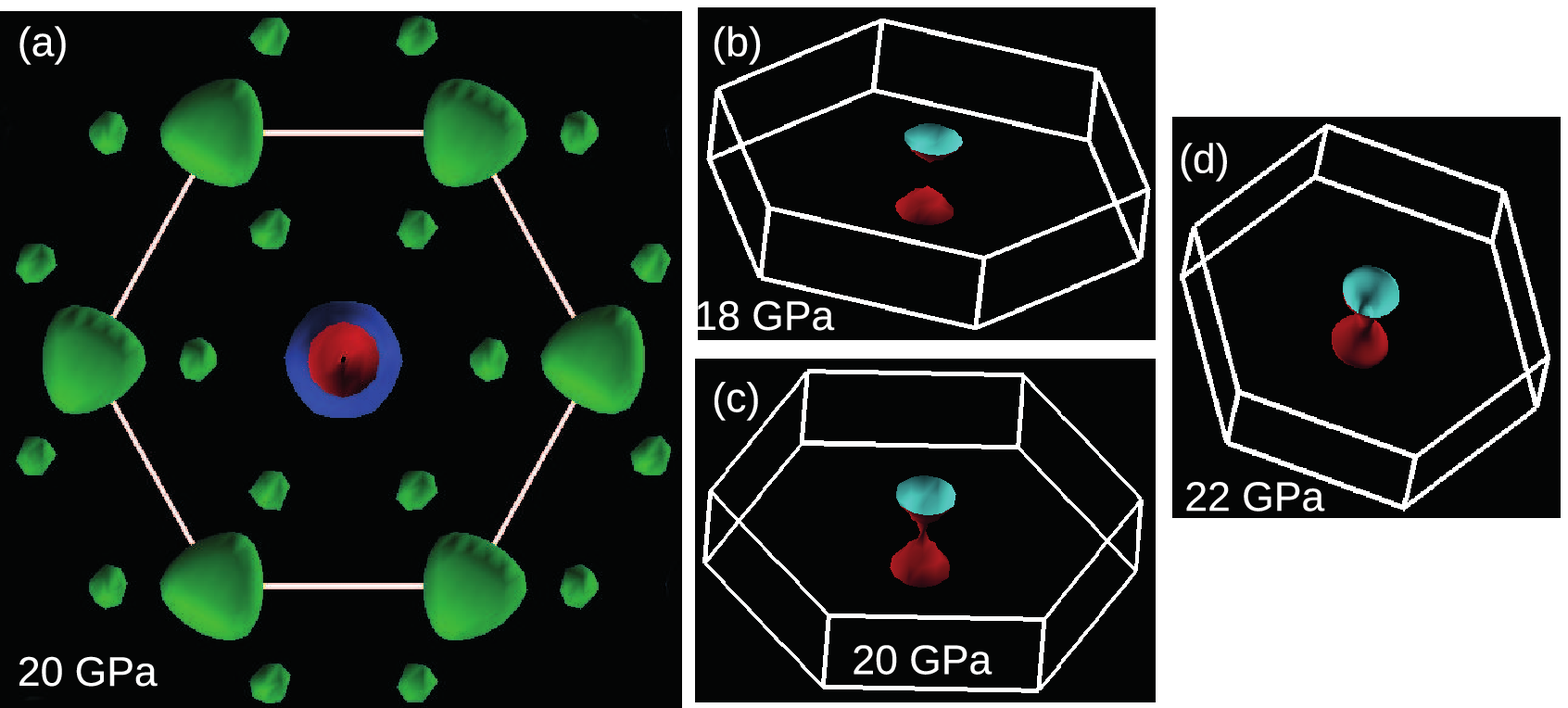}
	\caption{(color online)-- Fermi surfaces at (a) 20 GPa of merged bands (all the bands crossing the Fermi levels). Panels (b), (c) and (d) show the specific hole like part of Fermi surface changing with pressure at P=18 GPa, 20 GPa and 22 GPa respectively. Note that (b), (c) and (d) show the hole pockets at $\Gamma$ and A points with pressure. Green color shows electron pockets whereas blue and red (at the centre of the hexagon) show hole pockets.}
	\label{Fig_7}
\end{figure}

\par An increase in the pressure beyond 8 GPa creates new extrema in electronic dispersion with electron pockets (valleys) at K$_2$ and K points, and hole pockets at $\Gamma$ and A points (see Fig.~\ref{Fig_5}c). This emergence of hole and electron pockets seen clearly in Figs.~\ref{Fig_6}(a), (b) and (d) where the VBMs at $\Gamma$, K$_1$ and A points cross Fermi level at 10 GPa, 8 GPa and 12 GPa respectively, leading to formation of hole pockets. Similarly, CBMs at K$_2$ at 8 GPa and K points at 20 GPa cross Fermi level leading to the formation of electron pockets. As the hydrostatic pressure does not alter the symmetry of the crystal, energy levels do not split, but those near the Fermi energy change notably giving rise to pressure induced transfer of electrons from one pocket to another in order to maintain the number of carriers. Interestingly at 20 GPa, the gap at K-point closes (see Fig.~\ref{Fig_6}c), which within errors of our calculation, corresponds to the second transition experimentally seen at 16.5 GPa. To probe this further, we monitored the evolution of Fermi surface with pressure. At 20 GPa, we visualize Fermi surfaces associated with all the bands which cross Fermi level and find electron pockets at K-point and at K$_2$ along the path $\Gamma$ to K (see Fig.~\ref{Fig_7}a, green colour surfaces). In Fig.~\ref{Fig_7}a (red and blue surfaces), at $\Gamma$ and A points, we observe hole pockets (see Figs.~\ref{Fig_5}a and d). We find the Fermi surface associated with the bands at $\Gamma$ and A points changes significantly at an applied pressure of $\sim$ 20 GPa (see Figs.~\ref{Fig_7}b-d). Since, the Fermi surface changes with applied pressure without breaking the structural symmetry, we assign a Lifshitz transition at $\sim$ 20 GPa.

\begin{figure}[h!]
	\centering
	\includegraphics[trim=50 -20 15 -30, scale=0.8]{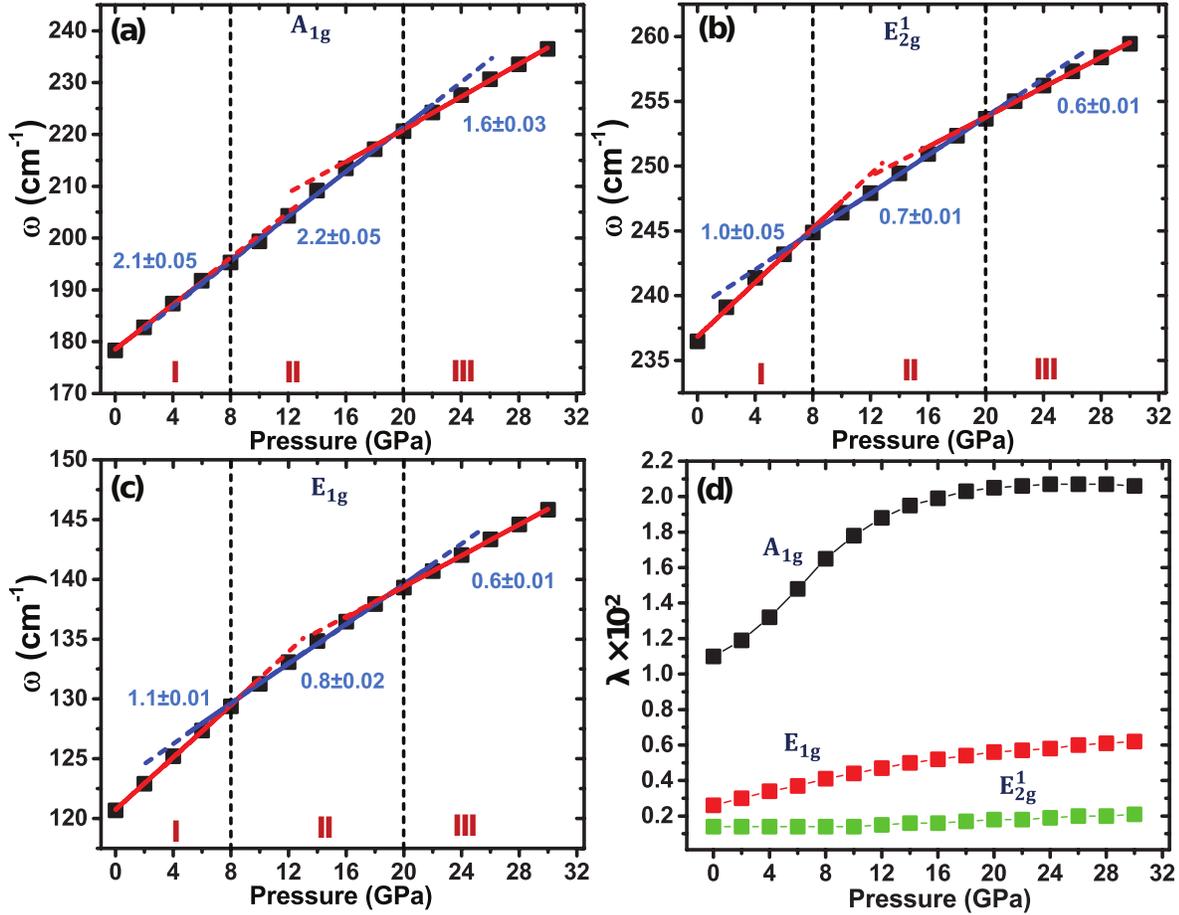}
	\caption{(color online)-- The pressure coefficients of Raman active phonon modes obtained using first-principles calculations. Changes in slopes (expressed in cm$^{-1}$/GPa) of A$_{1g}$, E$^{1}_{2g}$ and E$_{1g}$ are shown by vertical dashed lines in (a), (b) and (c) respectively. The changes in electron-phonon coupling of those modes are shown in (d).}
	\label{Fig_8}
\end{figure}

To investigate the pressure dependent phase transition from 2H to 1T$^{'}$ phases of MoTe$_2$, we study the changes in enthalpy of these structures, and did not observe any phase transition from 2H to 1T$^{'}$ phase. In fact, MoTe$_2$ in 2H form indicates increased stability with pressure (inset of Fig.~\ref{Fig_6}d). We investigated the phase transition only between 2H and 1T$^{'}$ phase as 1T$^{'}$ is the phase that is second lowest in energy. The energies of 1T$^{'}$-MoTe$_2$ and 1T-MoTe$_2$ are 133 meV/f.u. and 544 meV/f.u. with respect to 2H-MoTe$_2$, respectively. 1T-MoTe$_2$ is also locally unstable and exhibit structural instabilities with imaginary frequencies at K and M-points of Brillouin zone whereas MoTe$_2$ is stable in the 1T$^{'}$ form. The 1T$^{'}$ phase has a monoclinic lattice which is a distorted form of 1T phase. However, this structural distortion of 1T phase results in formation of weak in-plane metal-metal bonds in the pseudo-hexagonal layers with zigzag metal chains \cite{1t}.

We determined the effects of hydrostatic pressure on the Raman active modes. A compression of the unit cell leads to hardening of all the three modes A$_{1g}$, E$^{1}_{2g}$ and E$_{1g}$ (see Figs.~\ref{Fig_8}a, b and c). Here, the calculated pressure coefficients for all the Raman modes decrease after the SMT except for A$_{1g}$ mode, whereas the pressure coefficients for both the modes (A$_{1g}$ and E$^{1}_{2g}$) increase (see Fig.~\ref{Fig_8}a and Fig.~\ref{Fig_4}a) in experiments. This increase in the pressure coefficient may be further amplified possibly due to chalcogen vacancies present in the system which can influence its properties notably \cite{vac1,vac2,vac3,vac4}. While the difference in the magnitude of slopes of A$_{1g}$ of region I (0-6 GPa) and region II (6-16.5 GPa) in experiment is approximately 1 cm$^{-1}$/GPa (Fig.~\ref{Fig_4}a), it is underestimated in theory to be 0.1 cm$^{-1}$/GPa (Fig.~\ref{Fig_8}a). Similarly, we find difference in magnitudes of the calculated slopes of E$_{1g}$ and E$^{1}_{2g}$ from experimental slopes (see Fig.~\ref{Fig_4} and Fig.~\ref{Fig_8}). To explain this, we have examined the effects of anharmonic interactions between phonons. We froze A$_{1g}$ mode atomic displacements by 0.04 {\AA}, and determined the changes in E$^1_{2g}$ frequencies as a function of pressure. We find that the frequencies of E$^1_{2g}$ mode change by approximately 12-13 $cm^{-1}$, revealing that there is a relatively strong coupling between A$_{1g}$ and E$^1_{2g}$ modes. This anharmonic (phonon-phonon) coupling is not included in our analysis, may be responsible for difference in the slopes of A$_{1g}$ mode in region I and region II as a function of pressure in experiment and in theory.

\par We note that there are changes in slope (= $d\omega$/dP) of pressure dependence of all the Raman active phonon modes at 8 GPa and 20 GPa (From Fig.~\ref{Fig_8} (a, b and c)). A careful examination of the evolution of electronic structure with pressure indeed shows the pressure induced semiconductor to semimetal transition at 8 GPa and a Lifshitz transition at 20 GPa. Thus, there is a clear correlation between the slope changes of Raman active modes and electronic phase transitions, obtained within the same theoretical framework.

Fig.~\ref{Fig_8}(d) shows that the A$_{1g}$ mode couples more strongly with electrons than E$_{1g}$ and E$^{1}_{2g}$ modes (see Fig.~\ref{Fig_8}d). The size of electron pockets in the Fermi surface centered at K increases with pressure. This also can be understood with the help of group theoretical analysis of symmetry. The A$_{1g}$ mode has symmetry of the crystal (an identity representation). The electron phonon coupling (EPC) of A$_{1g}$ is large due to non-zero matrix element $\langle\psi_{\mbox{\scriptsize\boldmath $k + q$},i}|\triangle V_{\mbox{\scriptsize\boldmath $q$}\nu}|\psi_{\mbox{\scriptsize\boldmath $k$},j}\rangle$ \cite{epc1} for the perturbation A$_{1g}$ for all the electronic states. The EPC of A$_{1g}$ increases with pressure and gets saturated above 20 GPa (refer to Fig.~\ref{Fig_8}d). This saturation of EPC can be a result of the gap closing at K-point at 20 GPa. In contrast, E$_{1g}$ and E$^{1}_{2g}$ modes couple weakly as the matrix element vanishes for the E$_{1g}$ mode and is non-zero for a few of electronic states ($\it{i.e.}$ E$_{1g}$) for the E$^{1}_{2g}$ mode. Hence, we do not find any significant change in EPC with pressure for E$_{1g}$ and E$^{1}_{2g}$ modes.

Furthermore, to explain the non-monotonous change with a peak in relative intensity of A$_{1g}$ and $E^1_{2g}$ Raman modes (refer to Fig.~\ref{Fig_4}c) at the semiconductor to semimetal transition (P= 6 GPa), we estimated Raman tensors using first-principles calculations. Raman scattering intensity is proportional to square of Raman tensor and defined as,

\begin{equation}
 I \propto |e_{i}.R.e_{s}|^2,
\end{equation}

\noindent
where, $e_{i}$ ($e_{s}$) is the polarization of incident (scattered) radiation and R is the Raman tensor. Raman tensor is defined as,

\begin{equation}
R_{i\alpha\beta\gamma} = \frac{\partial Z^*_{i\alpha\beta}}{\partial E_\gamma}
= -\frac{\partial}{\partial u_{i\alpha}} \bigg(\frac{\partial^2 E_{tot}}{\partial E_\alpha \partial E_\beta} \bigg)
= -\frac{\partial \chi^\infty_{\alpha\beta}}{\partial u_{i\alpha}}
\end{equation}

\noindent
where,  $Z^*_{i\alpha\beta}$, $E_{tot}$ and $\chi^\infty_{\alpha\beta}$ are Born effective charges, total energy of the system and dielectric susceptibility (electronic contribution), and $E_\alpha$ is the applied electric field along $\alpha$ direction. u$_{i\alpha}$ is displacement of $i^{th}$ atom along $\alpha$ direction, and we use finite difference method to evaluate Raman tensor by freezing A$_{1g}$ and E$^1_{2g}$ modes with a magnitude $u_{A_{1g}, E^1_{2g}}$ (= $\pm$ 0.04 {\AA}), R = $\Delta{\chi}/\Delta{u}$.

\begin{table*}[h!]
\small
\centering
\caption{Components of Raman tensors of A$_{1g}$ and E$^1_{2g}$ modes, R (A$_{1g}$) and R (E$^1_{2g}$).}
\scalebox{0.8}{
\begin{tabular}[t]{c|c|c|c|c|c}
\hline
Pressure & R$_{11}$ (A$_{1g}$) & R$_{33}$ (A$_{1g}$) & $\sum_{i=1}^{2}$ R$_{ii}$ (E$^1_{2g}$) & $\sum_{i,j=1}^{2}$ R$_{ij}$ (E$^1_{2g}$) & Ratio\\
(GPa) & = R$_{22}$ (A$_{1g}$) & & = R$_{11}$ (E$^1_{2g}$) + R$_{22}$ (E$^1_{2g}$) &= R$_{12}$ (E$^1_{2g}$) + R$_{21}$ (E$^1_{2g}$) & $R_{11}(A_{1g})/R_{ij}(E^1_{2g})$\\
\hline
4 & 18.8 & 59.9 &0.0 & 4.4 & 4.27\\
8 & 23.5 &  131.1&0.0 & 0.4 & 58.75\\
12 & 27.7 & 514.7 & 0.0& 5.0 & 5.54\\
\hline
\end{tabular} }
\label{tab3}
\end{table*}

We find that R$_{33}$($A_{1g}$) increases with the pressure (see Table I), though its value after the gap closing point ($\sim$ 8 GPa) is not quite well-defined ($\it{i.e.}$ at P= 12 GPa). The R$_{ij}$ (E$^1_{2g}$) (= R$_{12}$ (E$^1_{2g}$) + R$_{21}$ (E$^1_{2g}$)) has a large magnitude at P = 4 GPa, and passes through a minimum at P = 8 GPa and then rises again. These elements do not change much above the pressure of the gap closing point. As a result, the relative intensity ratio of $A_{1g}$ to E$^1_{2g}$ modes will exhibit a maximum at 8 GPa. Thus, the peak in Fig.~\ref{Fig_4}(c) arises primarily from non-monotonous change in Raman tensor of E$^1_{2g}$mode.

\begin{figure}[h!]
   \centering
   \includegraphics[trim= 50 0 15 -20, scale=0.5]{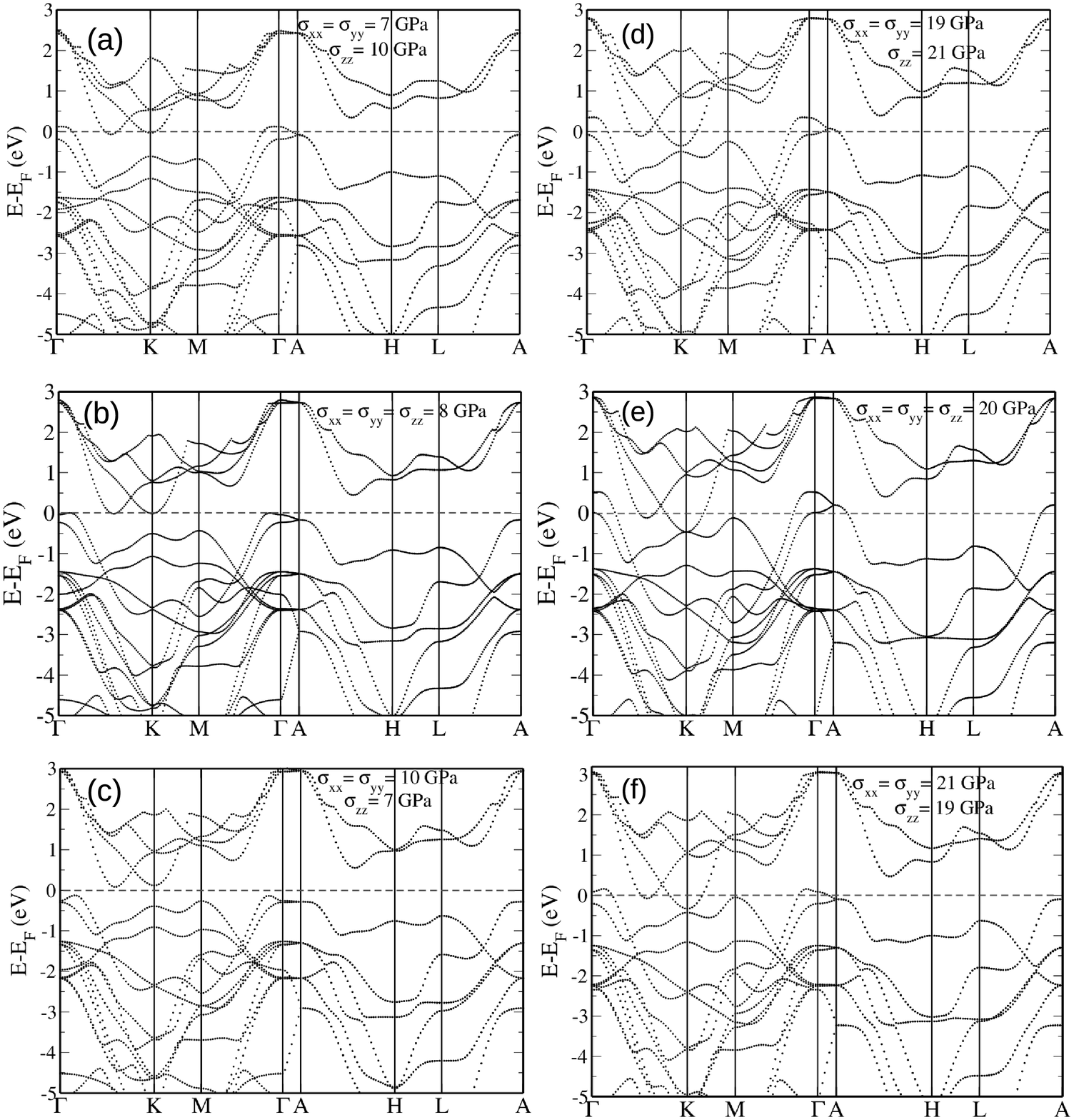}
       \caption{Electronic structure of 2H-MoTe$_2$ at non-hydrostatic pressure, (a) 
       $\sigma_{xx}$ (=$\sigma_{yy}$) = 7 GPa and $\sigma_{zz}$ = 10 GPa, (c) $\sigma_{xx}$ (=$\sigma_{yy}$) = 10 GPa and 
       $\sigma_{zz}$ = 7 GPa, (d) $\sigma_{xx}$ (= $\sigma_{yy}$) = 19 GPa and $\sigma_{zz}$ = 21 GPa, and (f) $\sigma_{xx}$ 
       (= $\sigma_{yy}$) = 19 GPa and $\sigma_{zz}$ = 21 GPa and at hydrostatic pressure (b) $\sigma_{xx}$ (= $\sigma_{yy}$ = 
       $\sigma_{zz}$) = 8 GPa and (e) $\sigma_{xx}$ (= $\sigma_{yy}$ = $\sigma_{zz}$) = 20 GPa.}
\label{Fig_9}
\end{figure}

It is known that the KBr pressure medium does not provide hydrostatic conditions above 3-5 GPa \cite{kbr1}. However, there are technical limitations in using other pressure transmitting media and hence we address this issue theoretically. For this, we compute the electronic structure of 2H-MoTe$_2$ at non-hydrostatic pressures near the transition pressures 8 GPa and 20 GPa, using first-principles calculations. At P = 8 GPa, 2H-MoTe$_2$ shows semiconductor to semimetal transition whereas Lifshitz transition takes place at P = 20 GPa. We consider two different non-hydrostatic conditions of pressure near 8 GPa ($\sigma_{xx}$ (=$\sigma_{yy}$), $\sigma_{zz}$) = (7, 10) and (10, 7) GPa. From our calculations, it is clear that 2H-MoTe$_2$ is semimetal at (7, 10) GPa (see Fig.~\ref{Fig_9}a) whereas it is semiconductor at (10, 7) GPa (see Fig.~\ref{Fig_9}c). Comparison of electronic structure of states at (8, 8) GPa and (7, 10) GPa reveals that density of states near Fermi energy is higher in the latter (Figs.~\ref{Fig_9}a and b). On the other hand, we find an opposite behavior at (10, 7) GPa, the density of states near Fermi energy decreases and a small gap opens up. Thus $\sigma_{zz}$ $>$ $\sigma_{xx} = \sigma_{yy}$ favors the transition at lower pressure, while $\sigma_{zz}$ $<$ $\sigma_{xx} = \sigma_{yy}$ pushes the transition to higher pressures. Thus, we conclude that (a) the character of the transition is preserved even when the pressure is non-hydrostatic, and (b) the transition pressure may change by a few GPa.

We perform similar calculations near second transition  (P= 20 GPa) at (19, 21) GPa and (21, 19) GPa (see Figs.~\ref{Fig_9}d, e and f), and find that small gap opens up at K point (Figs.~\ref{Fig_9}d and f), but it is well below the Fermi level. Clearly, there is no notable change in the states near Fermi energy. Thus, deviation from hydrostatic pressure should not affect the behavior of this higher pressure electronic transition, as much as it affects the lower pressure transition.

\section{Conclusions}
\label{sec:Conclusion}

In summary, we have analyzed the pressure induced semiconductor to semi-metal transition at $\sim$ 6 GPa and a Lifshitz  transition at $\sim$ 16.5 GPa in 2H-MoTe$_2$ by combining the Raman measurements and first-principles density functional theoretical calculations. The frequencies of the first order A$_{1g}$ and E$^{1}_{2g}$ Raman modes carry the signatures of semiconductor to semimetal and the Lifshitz transitions. The occurrence of a maximum in the integrated ratio of the A$_{1g}$ and E$^{1}_{2g}$ modes is mainly due to non-monotonous change in Raman tensor of E$^1_{2g}$ mode with pressure. We calculated the effect of pressure on Raman active modes, and find that pressure influences the EPC of A$_{1g}$ most strongly. All the Raman active modes harden with increasing pressure, and electron phonon coupling increases under compression due to changes in the Fermi surface. We hope that our findings will stimulate further study of high pressure and low temperature resistivity experiments to capture the anomalies near the Lifshitz transition. 

\section{Acknowledgments}

AKS acknowledges the funding from Department of Science and Technology, India. AKS and UVW acknowledge funding from a JC Bose National Fellowship. AB thanks CSIR for research fellowship. AS is thankful to Jawaharlal Nehru Centre for Advanced Scientific Research, India for research fellowship.

\end{document}